\documentstyle[12pt]{article}

\input {phyzmac.tex}
\input {refmac.tex}

\def\gtwid{\mathrel{\raise.3ex\hbox{$>$\kern-.75em\lower1ex\hbox{$\sim$}}}}
\def\ltwid{\mathrel{\raise.3ex\hbox{$<$\kern-.75em\lower1ex\hbox{$\sim$}}}}

\catcode`\@=11
\@addtoreset{equation}{section}

\catcode`@=12

\begin{document}

\begin{titlepage}
\begin{flushright}
BA-92-45
\end{flushright}
\begin{flushright}
IC-92-118
\end{flushright}
\begin{flushright}
{hepph@xxx/9207247}
\end{flushright}
\begin{center}
{\large{\bf Inflation and Large Scale Structure Formation After COBE}}
\vskip 0.1in
{\bf R.~K.~Schaefer and Q.~Shafi}\\
{\it Bartol Research Institute \\
University of Delaware, Newark, DE 19716\\}

\end{center}
\vskip 0.1in
\begin{abstract}

The simplest realizations of the new inflationary scenario
typically give rise to primordial density fluctuations which
deviate logarithmically from the scale free Harrison - Zeldovich
spectrum.  We consider a number of such examples and, in each
case we normalize the amplitude of the fluctuations with the
recent COBE measurement of the microwave background anisotropy.
The predictions for the bulk velocities as well as anisotropies
on smaller (~1-2 degrees) angular scales are compared with the
Harrison-Zeldovich case.  Deviations from the latter range from a
few to about 15 percent.  We also estimate the redshift beyond
which the quasars would not be expected to be seen.  The
inflationary quasar cutoff redshifts can vary by as much as 25\%
from the Harrison-Zeldovich case. We find that the inflationary
scenario provides a good starting point for a theory of large
scale structure in the universe provided the dark matter is a
combination of cold plus (~10-30 \%) hot components.
\end{abstract}
\begin{flushleft}
Pacs numbers: 98.80Bp, 98.80Cq, 98.50Sp
\end{flushleft}
\end{titlepage}

\section{Introduction}
\indent
   It was recently argued\CI{SS1} that the COBE
measurement\CI{COBE} of the large scale microwave background
anisotropy is in remarkable agreement with earlier
predictions\cite{r:SSS,r:Holtz} based on an inflationary
scenario\CI{Guth} in which the dominant mass component of the
universe is a mixture of cold plus hot dark matter\CI{SS}.  It
was tacitly assumed in these works that the primeval density
fluctuations resulting from the inflationary epoch take the
scale-invariant Harrison-Zeldovich form.

    It is well known, however, that the simplest inflationary
models lead to density spectra which deviate from the
Harrison-Zeldovich (hereafter H-Z) form through the presence of
logarithmic terms\cite{r:Linde,r:K+T}.  The precise forms of the
latter depend on the details of the model.  One of the
motivations for this paper is to rectify this situation.  The
deviations in some of the cosmological implications resulting
from this change are catalogued.  A new element in our analysis
is the normalization of the amplitude of density fluctuations
supplied through the COBE measurement.

The plan of the paper is as follows.  In section 2 we briefly
discuss some relevant aspects of the inflationary scenario and
the density spectra which arise in typical inflation models. We
consider the cosmological implications of these power spectra in
Section 3, with an emphasis on the normalization provided by the
recent measurement of the large scale microwave background
anisotropy by the COBE satellite.  We compute several important
quantities relevant for discussing large scale structure
including anisotropies on smaller ($1^\circ-2^\circ$) angular
scales, bulk velocities, and the bias parameter.  Attention is
paid to the very important issue of quasar formation at high
redshifts.  Some concluding remarks are given in Section 4.

\section{Density Fluctuations From Inflation}

   An inflationary scenario may be deemed successful if it
satisfies the following requirements:
\begin{enumerate}
\item {The duration $\tau$ of the inflationary phase is sufficiently
long to resolve the horizon and flatness problems,
\elab{tau}{\tau\gtwid 60\ H^{-1},}
where $H$ denotes the de Sitter Hubble constant
\elab{H}{H^2 = {8\pi\over 3}{\rho_v\over M_{p}^2},}
$\rho_v$ is the vacuum energy density and $M_p = 1.2\times 10^{19}$ GeV is
the Planck mass.}
\item  {The density fluctuations resulting from the quantum fluctuations
during inflation should respect the constraint imposed by the remarkable COBE
observations of large scale background anisotropy.}
\item  {After inflation a satisfactory mechanism must exist for generating the
observed baryon asymmetry of the universe.}
\end{enumerate}

    Reasonably successful inflationary scenarios are readily
constructed within the framework of ordinary and supersymmetric
grand unified theories.  (There exist several more elaborate
scenarios, some of which require an extension of Einstein's
general relativity, \eg\ Kaluza-Klein, extended inflation, \etc,
but we do not consider them in this work.)  The inflaton field is
typically a gauge singlet scalar field with suitably weak
couplings to ``ordinary" GUT matter.  The weak coupling
requirement is related to the small amplitude of the density
fluctuations generated through quantum fluctuations of the
inflaton.  This, in turn, imposes some restrictions on the reheat
temperature $T_r$ of the universe after the inflationary epoch.
One finds, typically, that $T_r\ltwid 10^{10}$ GeV.  Such low
$T_r$s were once dreaded, the fear being that it would be hard to
generate the baryon asymmetry.  This issue we believe has now
been put to rest with the realization\CI{L+S} that a most
convenient baryogenesis scenario within the inflationary
framework is obtained by first creating a lepton asymmetry
through the decays of the heavy right handed Majorana neutrinos.
(Recall that the hot dark matter is in the form of electron volt
mass neutrinos which, through the see-saw mechanism, requires
heavy Majorana neutrinos.)  The appearance of baryon and lepton
number violating sphaleron transitions at the electroweak scale
subsequently convert a fraction of the lepton asymmetry into the
observed baryon asymmetry.  This scenario for baryogenesis works
well both for supersymmetric and non-supersymmetric GUTs.

     The inflationary scenarios that we will consider lead to
density fluctuations which deviate from the pure H-Z scale free
density spectrum through the presence of the term
$[ln(\ell)]^\alpha$, where $\ell$ denotes the length scale and
$\alpha$ depends on the model.  The parameter $\alpha$ is
essentially determined from the quantity $[V(\phi)]^{3/2}/[M_p^3
({dV\over d\phi})]|_{k\sim H(\phi)}$, where $V(\phi)$ denotes the
potential energy density during inflation.  We will consider
three cases i) $\alpha=2$ (chaotic inflation with a quadratic
potential), ii)$\alpha = 3$ (quartic potential), and iii)
$\alpha=4$ (SUSY inflation).  Of course $\alpha=0$ corresponds to
the H-Z spectrum.

\section{ Cosmological Implications}

\indent
  We now proceed to compare the cosmological implications of
these inflationary density spectra with a pure scale-free (H-Z)
spectrum.  Inflation also predicts that the universe contains a
critical density of matter, which must be mostly non-baryonic if
we are to satisfy constraints from nucleosynthesis.  Inflation
does not predict what the dark matter is, however.  Usually it is
taken to be in the form of cold particles (cold dark matter or
CDM) which are relics of an earlier epoch in the universe.  The
other favorite candidate is a lightly massive neutrino, which
decouples from thermal equilibrium when relativistic or hot
(hence the name hot dark matter or HDM).  Nothing precludes a
mixture of the two types, indeed, a mixture is even expected in
certain grand unified theories.  A combination of about 10-30\%
HDM (and hence about 90-70\% CDM) has been shown to have
remarkable success at explaining some features of large scale
structure.  We take here the value 25\% HDM and 75\% CDM as our
canonical mixture for our model with cold plus hot dark matter
(or CPHDM).

  The procedure for fixing the amplitude of the spectra had
previously been made uncertain by the observation that the
distribution of light (bright galaxies) seems to be  different
from the distribution of mass.  This was treated by introducing a
density bias factor ($b$).  The fractional variance of the
galactic number  $N_{gal}$ has been measured to be 1 in a
randomly placed sphere of radius  8 $h^{-1}$ Mpc.  ($h$ is the
Hubble constant in units of 100 km s$^{-1}$ Mpc$^{-1}$; in this
work we will use the value $h=0.5$ to minimize conflicts with
constraints on the age of the universe.)  This is related to the
mass variance in the same spherical volume $via$ the bias
parameter $b$:
\elab{b}{\biggl\langle \left({\delta N_{gal}\over N_{gal}}\right)^2
(8h^{-1}\ {\rm Mpc})\biggr\rangle = 1 = b^2 \biggl\langle
\left({\delta M\over M}\right)^2 (8h^{-1}\ {\rm Mpc})\biggr\rangle.}
Many calculations have been performed to determine the appropriate value of
$b$ to account for the distribution of galaxies, with a range of $b\sim
1.5-3$ suggested for H-Z models when the dark matter comes in the
form known as cold dark matter (CDM).

    Since the remarkable measurement of primordial temperature
anisotropy by the COBE satellite\CI{COBE}, we can now normalize
the power spectrum independent of the value of $b$.  In fact, we
can use the COBE results to {\sl predict} the value of $b$ from
\eref{b}.  Implicit in our normalization to the COBE quadrupole
is the assumption that the anisotropy contribution from
gravitational waves\cite{r:rsv,r:AW1} is negligible.  (This is
the case for many realistic inflationary models.) We can also
predict various other features of large scale structure and
intermediate angle ($\sim 1^\circ$) temperature anisotropies.  We
begin by specifying our normalization technique.

   The quadrupole moment $Q_{rms}$ for a particular power
spectrum can be calculated from the formula\CI{AW2} involving an
integral over Fourier wavenumbers $k$
\elab{qrms}{ \left({Q_{rms}\over T_0}\right)^2 =
5\pi \epsilon_H^2 \int_0^\infty {dk\over k}
\left[T(k)\right]^2 \left[ j_2(kR_H) +
0.03 k {dj_2(kR_H)\over dk}\right]^2 [D(k)]^2}
where $T_0$ is the present photon temperature, $\epsilon_H$ is the
amplitude of the density fluctuations when they
cross the horizon, $T$ is the ``transfer function" which specifies the
relative linear growth of density fluctuations on different scales, $R_H =
5800\ h^{-1}$ Mpc is the distance to the last scattering surface,
 and $D(k)$ is
the function which represents the logarithmic deviations from a H-Z spectrum.
$D(k)$ is given by
\elab{D}{D(k) = \biggl[ 1 - {1 \over N_e}{\rm ln}(k\ 6000\ {\rm Mpc})
\biggr]^{\alpha/2}.}
where $N_e$ is the number of expansion e-foldings which had occurred
between the time the 3000 Mpc scale inflated larger than the
horizon and the end of inflation.
\eref{D} can be used directly to estimate the decrease in amplitude of
density fluctuations on scale $r\sim 1/k$
For example, in the quartic inflation potential, $N_e=66$ and $\alpha= 3$,
so at 8 $h^{-1}$ Mpc, the decrease is estimated to be $\sim 13$ \%.
Note that in the above formulae (\eref{D} and \eref{qrms}), setting
$\alpha=0$ recovers the pure H-Z spectrum result, for which
COBE found $Q_{rms}/T_0 = 0.61 \pm 0.15$.  The COBE analysis
explicitly used the H-Z (which they call $k^n$ with n=1.0) power
spectrum for their fit.  Their value would not change noticeably
for the inflation models we use here (\ie\ effectively n$\sim
0.97-0.94$). From the calculated quadrupoles, all models yielded
$\epsilon_H = (5.3 \pm 1.3) \times 10^{-6}$, with less than a 3\%
variation in the central value over the range of models
considered here.  This is because we have tailored the spectral
modification to have unit value on the horizon scale, where the
quadrupole is most sensitive.

  The COBE normalization has profound implications for smaller
scale structure.  To illustrate this, we plot in figure 1 the
contribution to $(\delta M/M)^2$ per logarithmic interval in $k$,
\ie $[d/d\ ln\ k][\delta M/M]^2$.  We have plotted this quantity
as realized both with CDM and our canonical CPHDM mixture.  For
large scales, $k<0.02$ Mpc$^{-1}$, the density fluctuations
depend only on the primordial spectra and not on the dynamics of
the dark matter.  In the CPHDM mixture the amplitude of density
fluctuations is reduced on small scales when a density
fluctuation is small enough that a HDM particle can escape it
within an expansion time.  The density spectra converge to a
single value on the horizon scale because of our normalization,
but diverge as we go to progressively smaller scales.  Thus we
expect the differences among the inflationary spectra to become
more pronounced as we proceed to smaller scales.  As we consider
different observational consequences in this section we will
start at the large scale end $1/k \sim 100$ Mpc of figure 1 and
systematically work our way down to the small scales at the right
hand edge of figure 1, \ie $1/k \sim 1$ Mpc.

   The COBE satellite measured temperature fluctuations on large
angular scales.  We would like to know what temperature
fluctuations one would expect on smaller angular scales.  The
calculation of smaller angle anisotropies is more complex than
for COBE angular sizes.  On the angular scale of $1^\circ$, which
corresponds to the $100h^{-1}$ Mpc comoving scale on the last
scattering surface, three main effects contribute to temperature
anisotropy: gravitational redshifts (the Sachs-Wolfe effect which
dominates the COBE signal), temperature fluctuations intrinsic to
the last scattering surface, and doppler shifts due to motions of
the last scattering matter(- see, \eg, \rref{peebyu}).  One
characterizes the resulting total temperature fluctuation by the
temperature fluctuation correlation function
\elab{C}{C(\theta,\sigma) = \biggl\langle {\delta T\over T}({\bf e_1})
{\delta T\over T}({\bf e_2})\biggr\rangle; {\bf e_1\cdot e_2} =
{\rm cos} (\theta).} In the above equation, the temperature
fluctuations are first convolved with a detector response
pattern, here taken to be Gaussian with width $\sigma$.  The
expected result of a single subtraction experiment, in which two
beams of width $\sigma$ and angular separation $\theta$ are
differenced, can be represented by the formula\CI{WS}
\elab{dT}{\Delta_{rms} = \Bigl[ 2 C(0,\sigma)-2 C(\theta,
\sigma)\Bigr]^{1/2}.}
Sensitive experiments to search for these temperature
anisotropies have been performed in Antarctica with the most
sensitive of the published results\CI{Mein1} being $\delta T/T
<3.5\times 10^{-5}$ on an angular scale of 1$^\circ$ ($\sigma =
0.22^\circ$).  In Table 1 we report the predicted results of this
experiment for all of the models under consideration.  With the
COBE normalization, the temperature anisotropies are not very
sensitive to the type of dark matter.  They are however,
sensitive to the baryonic fraction, here taken to be the central
prediction of big bang nucleosynthesis for $h=0.5$, \ie,
$\Omega_b = 0.05$.  The results scale roughly as $(\Omega_b
/0.05)^{0.1}$.

   While all of the models seem to be quite compatible with the
95\% confidence limit at 1$^\circ$, there is a preliminary
report\CI{lubin} of a limit at 2.1$^\circ$, ($\delta T/T <1.4
\times 10^{-5}$, $\sigma = 0.63^\circ$), which  seems to be
somewhat in conflict with inflationary predictions.  For CDM
models with the COBE normalization we predict $\delta T/T$ of
1.76, 1.68, 1.65, and 1.61 for the H-Z, chaotic, quartic and SUSY
models, respectively. Using a mixture of cold plus hot dark
matter does not change these values.  This is because at the
length scales as large as those corresponding to $2^\circ$,
there is virtually no difference in the spectra as displayed in
figure 1.  One must remember that the COBE result, which
normalizes these predictions, has formal errors of $\pm25$\%, so
the predicted anisotropy is compatible with this preliminary
result within errors.  One may note that  the conflict with this
datum is eased with inflationary spectra, albeit not by much.

   On a slightly smaller scale, large patterns of galactic flow
have been observed and sophisticated analysis of these
observations has been done to characterize their velocities.  For
example, the three dimensional bulk motions of galaxies have been
derived\CI{Mred} by constructing the self-consistent velocity
field from galactic motions along the line of sight and assuming
that all velocities are induced by the action of gravity.  The
density field is first smoothed on a large scale, $R_s= 12
h^{-1}$ Mpc, to minimize the noise.  The average bulk velocity
within spheres of radii 40 $h^{-1}$ Mpc and 60 $h^{-1}$ Mpc were
then calculated and the results are given at the bottom of Table
1.  We compare the predictions of the inflationary models
calculated by the formula
\elab{V}{ \Bigl\langle V^2 (R)\Bigr\rangle =
4 \pi \epsilon_H^2 \left( {c\over H_0}\right)^2 \int dk\ k\
W^2(kR) e^{-k^2 R_s^2}\left[T(k)\right]^2 [D(k)]^2,} and the
results are given in Table 1.  As has been noted
previously\cite{r:Gorski,r:goodguys}, the COBE implied
normalization of the scale free spectrum is in remarkable
agreement with the large scale bulk motions.  The inflationary
logarithmic terms degrade the agreement somewhat, but are still
consistent (within errors) with the observationally derived
values.   We note that the Bertschinger
\etal\ [\rref{Mred}] velocities are also consistent with maximal likelihood
estimates from Faber (see, \eg, \rref{Holtz} or \rref{goodguys}).

   As we proceed to smaller scales we come back to the issue of
galactic density biasing.  We now calculate the bias factor, $b$
using \eref{b} above and
\elab{dM}{ \Bigl\langle \left({\delta M\over M}\right)^2 (R)\Bigr\rangle =
4 \pi \epsilon_H^2 \left( {c\over H_0}\right)^4 \int dk k^3
W^2(kR)
\left[T(k)\right]^2 [D(k)]^2,}
where $W(kR)$ is the ``window function", \ie\ the Fourier
transform of a sharp edged sphere of radius $R$ (see \eg,
\rref{Peebook}).   We note that figure 1 is just the integrand of
\eref{dM} with the window function removed.  We first present the
bias factor results for the CDM case in Table 1 using the
inflationary models described in section 2.  The first point to
note is that in all cases CDM bias factors are $\ltwid 1$.  Such
values are problematic for CDM.  Most N-body simulations favor
$b\sim 2-3$ to match the observed galactic properties
\cite{r:gelb,r:FWED}.  With $b>1$, one also expects virial
estimates of galactic cluster masses to underestimate the true
cluster mass because the light is more clustered than the matter.
Since present virial cluster mass estimates typically find only
about 20-30\% of the critical density, $b>1$ gives a convenient
explanation for how the observations can be reconciled with a
critical density universe.  The precise amount of biasing
required to explain virial mass estimates cannot be specified
because there are other dynamical uncertainties, \eg\ the
relative importance\CI{bv} of ``velocity biasing".  However, the
COBE implied amplitude is a serious problem for CDM models.  One
way to reduce the amplitude of density fluctuations on small
scales is to replace some of the CDM with hot dark matter.  In
figure 1 we have also plotted spectra for inflationary models for
our canonical CPHDM mixture (25 \% HDM and 75 \% CDM)) as is
favored by large scale structure
observations\cite{r:goodguys,r:Holtz,r:SSS}.  We have calculated
the bias parameters for these CPHDM models in Table 1 and they
are $\sim 1.3-1.4$, close to predictions\CI{goodguys}.  Galaxy
formation in these models has not been as extensively studied as
in CDM, although preliminary results look
encouraging\cite{r:MDM,r:gguys2}.  This is also marginally
consistent with large scale determinations of the bias
parameter\CI{Mred} which find $b<1.3$.

    As a last test of structure formation, we consider the
formation of high redshift quasars.  Since those objects
correspond to galactic scale density perturbations, which appear
at the right edge of figure 1, we expect the inflationary spectra
to display differences larger than in any other effect considered
in this work.

   Quasars are assumed to be extremely massive black holes
surrounded by infalling radiating matter.  The black holes would
be the result of the gravitational collapse of density
perturbations generated from inflation.  Since quasars are
observed at very early times (high redshifts), this gives a
constraint on the amplitude of density fluctuations on small
length scales.  Efstathiou and Rees (1988) [hereafter, ER] have
addressed this issue in the context of a H-Z spectrum with cold
dark matter.  Their approach can be summarized as follows.  They
estimate the size of the quasar central black hole, based on the
quasar luminosities.  Despite the presence of large amounts of
dark matter, this black hole is presumably composed mostly of
baryonic matter, because the non-baryonic matter is taken to be
dissipationless, and has difficulty shedding the angular momentum
required for collapse to a compact object.  Since  the baryons
are only a small fraction of the matter in the universe, one
needs a total mass fluctuation which is considerably larger than
the central black hole mass.  The black hole must be contained
within a  fluctuation with a mass at least as big as
\elab{halo}{M \approx 2\times 10^{12} f M_\odot} in order to
insure a black hole large enough to generate a luminosity of
$\gtwid 10^{47}$ ergs/sec, where $M_\odot = 2 \times 10^{33}$ gm
and f is a ``fudge factor" which incorporates uncertainties about
the quasar lifetime, the radiation efficiency and the fraction of
matter in the density perturbation which comprises the central
black hole.  $f$ is unlikely to be $<1$.

   ER estimated the number density of quasars at a redshift of 2
to be $1.2\times 10^{-7}$ Mpc$^{-3}$.  They estimated the number
of potential CDM quasar halos using the Press-Schecter theory.
Assuming every such halo forms a quasar, the number $N$ of
quasars with luminosity greater than $10^{47}$ ergs/sec at a
redshift $z$, and lifetime of $10^8$ years, can be found from
\elab{nq}{N(z) \simeq 1.8 \times 10^{-3} (1+z)^{5/2}{1\over \sigma_0}
{\rm exp}\biggl(-{(1.33)^2(1+z)^2\over 2 \sigma_0^2}\biggr) {\rm
Mpc}^{-3}} where $\sigma_0$ is the rms amplitude of the mass
fluctuation on the scale of the minimum mass halo which could
give rise to a quasar (and can be calculated from \eref{dM} with
the substitution of a Gaussian window function $W={\rm
exp}[-k^2R^2]$).  Since every halo will not necessarily produce a
quasar, \eref{nq} represents the maximum possible quasar density
at redshift $z$.  One can see that at some redshift there will be
an exponential cutoff in the allowed number density of quasars.
ER used a H-Z density spectrum with strongly biased (b=2.5) cold
dark matter in
\eref{nq} and found that the number density of quasars must drop below the
z=2 density at redshifts of $\gtwid 5$.  The quasar density must
decrease beyond that redshift.  We will refer to the redshift at
which the quasar density drops below the estimated z=2 density as
the {\sl cutoff redshift}, $z_c$.  The discovery of many (20 with
redshifts between 4 and 5 - \rref{qs}) quasars at redshifts
greater than 4 caused some to wonder about the viability of a H-Z
spectrum with CDM based on the ER calculation.  The spectra we
consider here contain less power for making high redshift quasars
compared to a similar H-Z spectrum.  However, ER used a high bias
factor, $b\sim2.5$, which was derived from H-Z predictions of CDM
galactic properties, and cautioned that their results are quite
sensitive to the normalization.  The redshift $z_c$ scales with
the normalization as
\elab{z}{(1+ z_c^\prime) = {[\delta M/M (R_q)]^\prime\over [\delta M/M (R_q)]}
(1+z_c)}
where $R_q = 1.2$ Mpc is the radius of a sphere which encloses
the minimum quasar halo mass.   With the bias factor implied by
the COBE result, $b=0.8$,  the redshift for the onset of quasar
formation gets scaled back considerably to $z_c\sim (2.5/0.8)
(1+5) -1 = 19 $, an incredibly high value, indicative of the
overabundance of galactic scale power in low-bias CDM.  We have
used \eref{nq} to calculate more carefully  the cutoff redshifts
in the inflationary CDM models and the cold plus hot dark matter
models and presented the results in the final column of Table 1.

     The COBE implied normalizations of CDM inflationary models
show a range of $z_c$ (quasar cutoff redshifts) from 15.8 for
SUSY inflation to 17.5 for chaotic inflation as compared with
19.9 for a pure H-Z spectrum.  There is thus no danger of
conflict at least with high redshift quasar observations.

     In the CPHDM models (with 25\% HDM) the cutoff redshifts are
much closer to observed quasars redshifts.  At constant bias
factor, the amplitude of mass  fluctuations on quasar scales is
smaller than the CDM value by a factor of about 1/1.7 (the same
amplitude as a $b=1.7$ CDM model).  The equivalent quasar
redshift for the H-Z spectrum and the COBE normalization for
CPHDM is $z_c = 6.5$.  From Table 1 we see again that adding the
inflationary logarithmic terms reduces $z_c$, although all
$z_c\geq 5$.  These latter values should soon be testable.
However, the numbers are sensitive to the quasar scale and the
normalization, so we do not base firm conclusions on these
estimates.  More detailed comparisons of this model for quasar
formation is under way\CI{gguys2}.  The COBE derived amplitude
uncertainties can also alter $z_c$ by 25\%.

\section{Conclusions}

   We have looked at cosmological structure formation models with
more realistic inflationary density fluctuation spectra and
compared them to the Harrison-Zeldovich spectrum.  For the
non-baryonic dark matter required by inflation and big bang
nucleosynthesis we have used both pure cold dark matter and a
mixture of 1/4 hot dark matter and 3/4 cold dark matter.  The
amplitude of the density fluctuations was normalized to the COBE
measurements.

    With this COBE prescription, we find that galactic scale CDM
density fluctuations are too large, as symptomized by the
implication that matter must be more  clustered than light: bias
factor $b<1$.  The inflationary logarithmic contributions to the
density fluctuation spectrum lessen the amount of galactic scale
power, but not by enough to escape the problem.

   Models with a mixture of cold plus hot dark matter (here we
have explicitly studied 75\% CDM $+$ 25\% HDM) seem to be much
more compatible with the requirements  for galaxy formation.  In
this case, the COBE normalization of the density fluctuation
amplitude implies a bias parameter compatible with observations
of clusters of galaxies.  The smaller amount of galactic scale
power implies a sharp drop in quasar density for redshifts
$\gtwid 5$ ($\gtwid 6$ for a pure Harrison-Zeldovich spectrum).
Thus the  density fluctuation amplitude in CPHDM models seems to
be compatible with  that implied by the COBE central value, but
cannot vary much from this  value.  Increasing the amplitude
could cause a serious conflict with  the 2.1$^\circ$ temperature
anisotropy (assuming the preliminary result does not change).
Decreasing the amplitude could cause conflict with observations
of high redshift quasars.   One may tentatively conclude,
however, that an inflationary scenario  with dark matter
composition consisting of 10-30\% HDM, the rest being CDM and
baryons, provides a good starting point for a theory of large
scale structure formation.

\section{Acknowledgement}
We would like to thank NASA and DOE for support.  Q.S. would also
like to thank the High Energy Physics group and Prof. Abdus Salam
at the ICTP for their hospitality.

\begin{table}
\begin{center}
{\small
\begin{tabular}{||l|c|c|c|c|c|c||} \hline\hline
\multicolumn{7}{||c||} {Measures of Cosmological Large Scale Structure} \\

 Spectrum &Dark &$\Delta_{rms}(1^\circ)$ &V(80 Mpc)&V(120 Mpc)&$ b$&$z_c$ \\
        &Matter &$\times 10^5$ &km/s & km/s & & \\ \hline
 & & & & & &  \\
 Harrison-  &CDM  &  2.68  & 387 & 307 & 0.80 & 19.9 \\
 Zeldovich   &CPHDM & 2.68 & 383 & 308 & 1.27 & 6.5 \\
 & & & & & &  \\
Chaotic  & CDM  & 2.57 & 361 & 288 & 0.88 & 17.5 \\
 Inflation  & CPHDM & 2.56 & 358 & 289 & 1.39 & 5.6 \\
 & & & & & &  \\
Quartic & CDM      & 2.50  & 356 & 284 & 0.90  & 16.8 \\
 Inflation & CPHDM & 2.50  & 353 & 285 & 1.42  & 5.4 \\
 & & & & & &  \\
SUSY       &  CDM  & 2.45  & 345  & 276  & 0.94  & 15.8 \\
 Inflation & CPHDM & 2.44  & 342  & 277  & 1.46  & 5.1 \\
 & & & & & &  \\    \hline
Observed &         & $<3.5$ & $388\pm67$ &$327\pm82$&&$>4$\\
\hline
\hline
\end{tabular}
}
\caption{All predicted values are subject to the COBE normalization
uncertainty, \ie, $\sim$25\%. Note, however, that the bias parameter $b$
varies inversely with the COBE quadrupole amplitude, unlike the other
parameters in the table.}
\end{center}
\end{table}
\vfil

Figure 1.  Mass fluctuation squared per logarithmic interval in
Fourier ($k$) space.  The top set of curves are for inflationary
fluctuation spectra evolved with CDM, while the bottom curves are
evolved with a mixture of 75\% CDM plus 25\%HDM.  The dark matter
dynamics are unimportant on scales larger than $1/k\sim 50$ Mpc.
All of the curves will converge at $1/k\sim 3000$ Mpc because we
have normalized all spectra to the COBE measurement.

\end{document}